# SEMANTIC INTEGRATION PROCESS OF BUSINESS COMPONENTS TO SUPPORT INFORMATION SYSTEM DESIGNERS

Hicham Elasri and Abderrahim Sekkaki

Departement of Mathematics and Computer Science University
Hassan II, Ain Chock, Faculty of Sciences Casablanca, Morocco

hicham_elasri@yahoo.com a.sekkaki@fsac.ac.ma

***ABSTRACT***

*The present work is inscribed within the intersection of two scientific thematic: the engineering by reuse of components and ontologies alignment. The integration of Business Components (BC) is a research problem that has been identified in the field of engineering by reuse. Our proposal aims to provide assistance to designers of information systems in the integration phase. It is a process guided by domain ontology to provide semantic integration of BC. This process allows the detection and resolution of semantic conflicts naming type encountered in the process of integration of BC.*

## 1. INTRODUCTION

Industries must increasingly face the constraints of cost, time and effort, the latter are heavily invested in activities and complex tasks from trades companies, One of the most complex activities is modeling a business in order to building an information system. In fact reuse is a sensible approach because of its way to address these constraints. Reuse of domain knowledge, especially those from a particular reuse of component for developing new business Information Systems (IS) from reusable components, this last approach well-known in *design by reuse*, today is widely adopted and used [1], [2], [3]. Using this approach includes implementation phases as well as preliminary phases of analysis and design. However, components needed during design and analysis phases are not technical but conceptual. In fact, this class of Components implements business logic and knowledge of a domain. Components involved in analysis and design phases are commonly referred to as Business Components (BC). In recent years, many different approaches focused on how to design new IS from reusable components [1], [2]. Two ways of research in the area of the reuse are intensively explored. The first one called “design for reuse” is to develop methods and tools to produce reusable components. The second “design by reuse” is to develop methods and tools to exploit reusable components [4]. We are concerned in this research by the second way. Literature outlines several questions when we address the topic of designing a new Information system by reusing available components. The main problem during development of information system is to ensure an effective reuse. This is why; it appropriate and necessary to predict an integration activity which includes a set of BC into one. In fact, Integrating into the same IS of several business components which emanate from various sources produces different conflicts both syntactic and semantic. We focus in this work on detecting and resolving semantic name conflicts encountered during the integration process of business components [5], [6] and [7]. We assume that the design of an IS intended generally a business domain and that business components model fragments of this domain. Otherwise, semantic integration systems are mostly based on the alignment of ontologies; this issue has given rise to several works [16], [5] and [6]. We relied on results of these works to support semantic integration process and have proposed semantic integration process based on the alignment of ontologies using domain ontology and a method of measuring semantic similarity. However, this solution allows creating semantic relations between concepts that may generate conflicts, but does not present how to use this relationship as to achieve semantic integration. To overcome this insufficiency, we propose an extension of our semantic process integration [6], in the present work using rules derived from semantics relations detected in semantic matching process in order to generate actions for resolving conflicts for propose to information system designers. We will validate our results using a prototype that we have developed and tested on domain ontology and some BC. Our paper is organized as follow: a proposal of business component meta-model is presented in section 2, semantic integration process of business component are described in Section 3. In section 4 an example of application and a prototype are presented in order to illustrate our proposal. In In section 5 a discussion on the use of isomorphic graphs (ontology) to improve consisting of our integration process. Finally, section 7 presents the conclusion and perspectives of our work.

## 2. BUSINESS COMPONENTS METAMODEL

Business Component (BC) aims to reduce significantly costs and cycle-time of developing software, time of maintenance and risk. Components based approach consists in building new systems by reusing available components. Using this approach in the earliest phases of system development presents a real interest. According to this approach, a business IS will be built from a set of BC which are generally heterogeneous. In fact, these BC generally emanate from various sources. For example, a company trading IS could be designed from multiple BC such as: {"Sales", "Product", "Customer» etc...}.

In order to realize an integration of business components, we need a set of common standards and language for BC. In our context, the BC candidates for integrations are described with heterogeneous languages and for integrate them we need to transform the presentation languages of BC to a common language based on the MDA approach in this sense, we propose a hybrid business components Meta-Model based on Meta-Model proposed in [8], [9] and [10]. The underlying motivation for metamodeling within the context of MDA is analyzed in [11].

Herzum and Sims in [9] state that business components realize business processes or business entities. While the Meta model by Herzum and Sims defines an essential two types: entity and process components and [9] state that reusable business components define a unique structure for a business object. This structure is reusable in any context. Those components are called —reusable because they can be reused —simply by integration of the proposed structure within a conceptual schema. [10] Add a class of business component called generic business components: those components define several different structures for a same business object; based on those approaches we propose the Meta model of business component (see Figure 1)

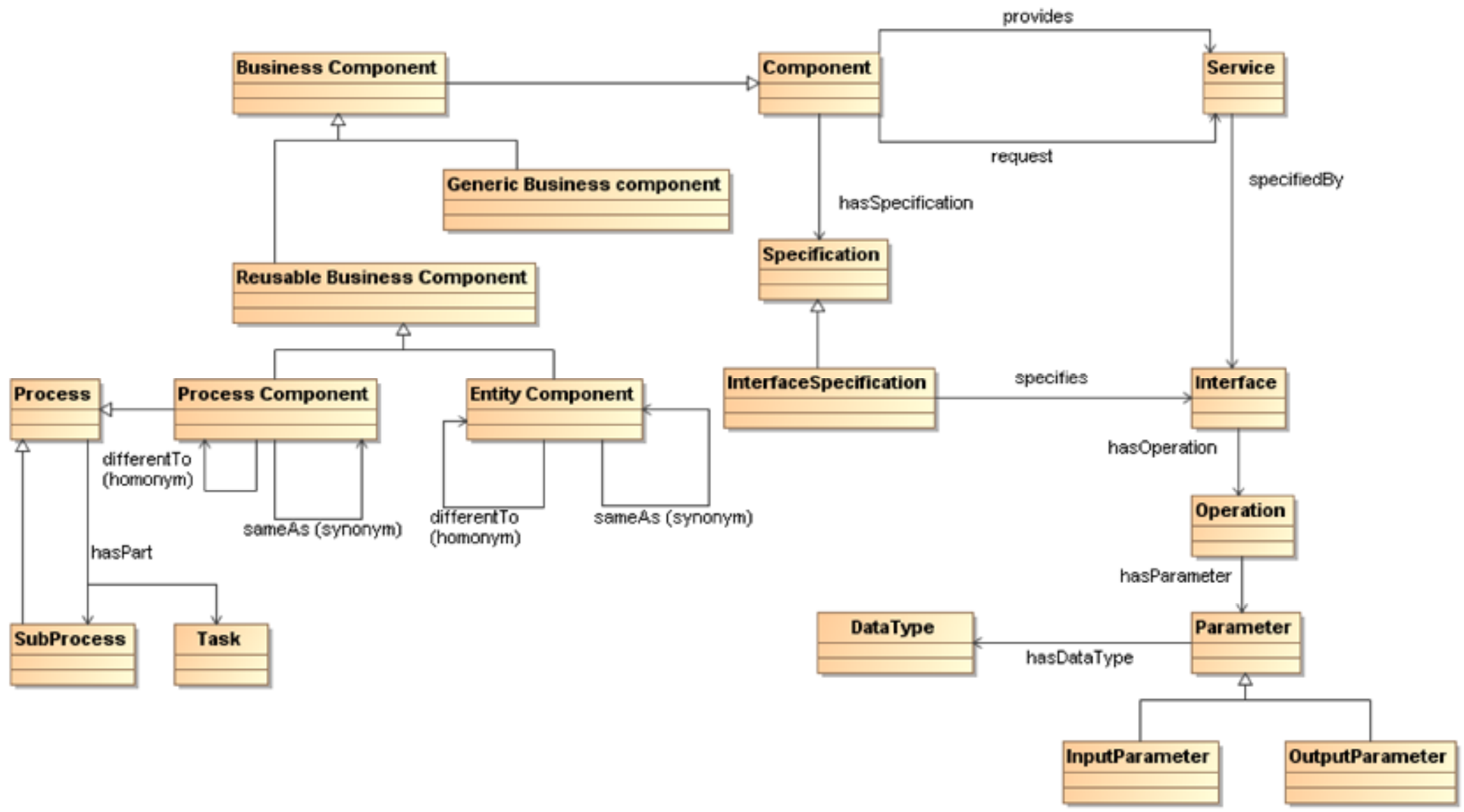

Figure 1. Business Component Meta-Model

## 3. SEMANTIC INTEGRATIONOF BUSINESS COMPONENTS.

The semantic integration of different BC in the same SI goes through a process of detection and resolution of semantic conflicts that may exist between different components. We believe that every conflict is generated by a non-definition of a semantic relation (eg synonymy semantic relationship which may cause a conflict type naming).We based in this work on the alignment of ontologies to align the ontologies associated with BC. Because of its ability to produce what we call Correspondences Ontology (CO) which includes the concepts and their semantic relationships from multiple sources ontologies. This task required and appropriate in the process of semantic integration, this is how we show the usefulness of CO and see how it can be used either in an automatic process as input of the phase integration is a process assisted by the designers of SI. Thereby deducting a set of actions (add, edit or delete a concept or relation) in order to achieve semantic integration of BC.

The integration of BC aims to detect and resolve conflicts caused by the heterogeneity of BC. The goal is to produce a single unified component. Moreover, components to integrate describe fragments of business knowledge in a language chosen by their designers. Several studies have focused on the transformation of BC described in modelling languages such as UML to ontologies. We have proposed in [5] and [6] integration processes that reduces the problem of semantic integration of BC to a problem of ontologies alignment. We are based on the definition proposed in [12] to define integration of business components: The integration of business components takes a set of components: BC1… BCn and correspondence model C1….n between them as input and combines their elements into a new output component BC1.... n.

$$BC1....n = Integration\ (BC1,...\ BCn,\ C1....n)$$

The semantic integration of BC takes a set of components: BC1… BCn and correspondence model C1….n which can be a correspondence ontology between them as input and combines their elements into a new output component BC1.... n., which means:

$$CO1....n = semanticIntegration\ (BC1... BCn, CO1....n)$$

We use domain ontologies for multiple reasons: Firstly, domain ontologies describe concepts related to a domain, this corresponds fully with our problem, since the design of an IS intended generally a business domain. Secondly, domain ontologies are reusable inside the same domain [13], this property is very interesting to consider in BC reusing, which is the central aim of design by reuse approach.

## 3.1. Ontologies Alignment

Ontologies are recently initiated approach for structuring knowledge and are defined as a collection of concepts, their interrelationships, axioms and proprieties which provide an abstract view of an application domain. According to Gruber, ontology is defined as an explicit formal specification of terms of a domain and relations among them [14].

It appears increasingly necessary to be able to reason on ontologies. To assess and align or match them in the perspective of solving problems of understanding and interpretation of the data.
*A matching process can be seen as a function f which tackes two ontologies O and O'', a set of parameters p and a set of oracles and resources r, and retunrns an alignment A between O and O''* [15].

The alignment process takes as input two ontologies O and O ', a set of parameters p and a set of resources r and provide into output an alignment between O and O'. Aligning ontologies consists in establishing semantic relations among concepts of various ontologies which describe the same field of knowledge. Aligning ontologies represents a great interest in application domains that manipulate heterogeneous knowledge.

*"Using comprehensive background knowledge in form of ontology can boost the ontology matching process as compared to a direct matching of the two ontologies."*

Several works on the alignment of ontologies have emerged over recent years; most of them are based on an external resource that can be either a general ontology or domain ontology [16], [17]. Similar experiments with similar results are described in [18]. The use of textual and lexical resources, in particular the case of WordNet as knowledge support or background, what was proposed by many researchers [15]. [19] [20], [16]. An original proposal comes from [21] which analyzes the semantic resource available online.

## 3.2. Business Component Integration Process.

Business Components provide services and / or data which are expressed in most cases, in a terminology freely chosen by their designers. Semantic integration of BC consists to attribute meaning to data and services in order to ensure their integration among heterogeneous BC and thus to allow their integration into the same IS. We propose in this section an extension of the solution that we have presented previously in [5], [6]. Our solution allows:

- Detection and resolution of semantic name conflicts among components business to integrate into the new IS.
- Production a new BC obtained from the integration of original business components.
- Propose guidelines or rules derived from the integration of a set of relationships matches.

This solution consists of two complementary sub-processes:

- *The process of semantic pre-integration.*
- *The process of semantic integration.*

A global description is provided in the following figure:

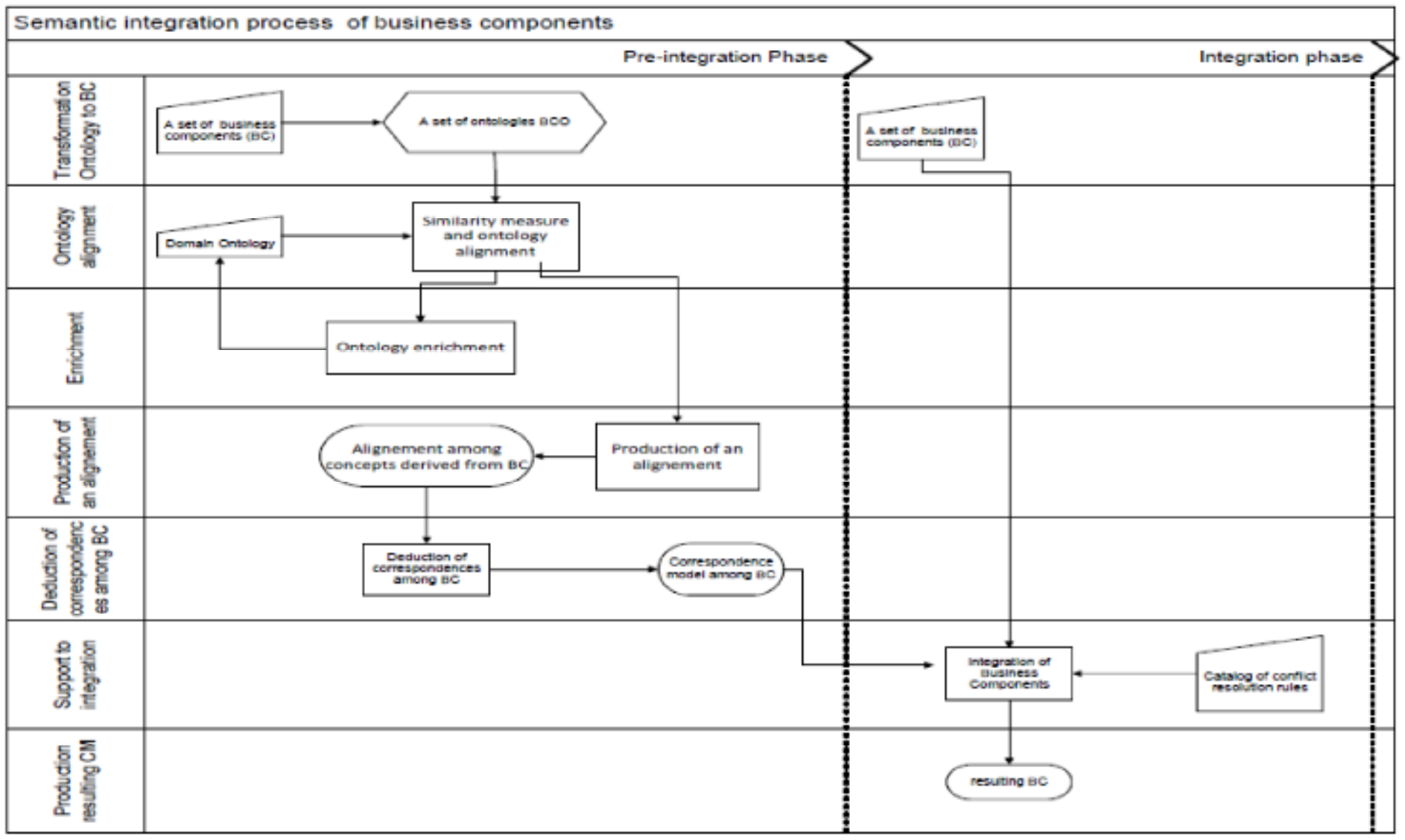

Figure 2. Global view of Business Component integration Process

### 3.2.1. The process of semantic pre-integration.

The aim of this process is the production a set of semantic relation between concepts derived from the BC candidates for integration, represented by a correspondence ontology. This process consists of a process description is provided in the following:

The inputs of the integration process are:
- A set of Business Components selected by the designer in order to integrate them in the future information system. We denote BC1….BCn, these BC.
- A domain ontology chosen by the designer according to the new IS domain. The domain ontology describes concepts and relations among concepts of the IS domain.

One output obtained at the end of the integration process:
- Correspondence ontology (Alignement): In the first step, IS designer can use this ontology to detect and resolve semantic conflicts in a semi-automatic process. In the second step, the ontology could be reused in an automated process from the perspective of integrating BC while defining a set of integration rules derived from the correspondence of BC. It will later be used as ontology support during the second process: the integration process.

An correspondence ontology can be used as entry the integration process and can be used to update the original domain ontology.

The pre-integration process comprises the following steps:

1. Transformation the BC candidates for integration into ontologies
2. Aligning ontologies obtained based on background ontology.
3. Produce correspondence ontology.

*A. Business Component transformation into ontologies.*

Several research studies have focused recently on the transformation of conceptual models described in a language such as UML into models using ontology description languages such as OWL. Thus [22] proposes a model driven (MDA) based methodology to generate ontologies from an annotated UML business model. Gasevic works [23] allow generating ontology from an UML model annotated by UML profile stereotypes of OWL provided by ODM (Ontology Definition Model). Transformations are performed by XSLT style sheet applied on XMI format models. A comparison between models and ontologies is given in [25]. The differences between the classes of the UML and OWL are studied in [26] and [27]. [28] provided an analysis of approaches for transforming UML to ontologies and another approach for transforming UML to OWL2 have been presented in [24]. This transformation is shown below.

Relying on the results of these studies, each BC candidate for integration is transformed into ontology, thus bringing the problem of BC semantic integration to a problem of ontology alignment.

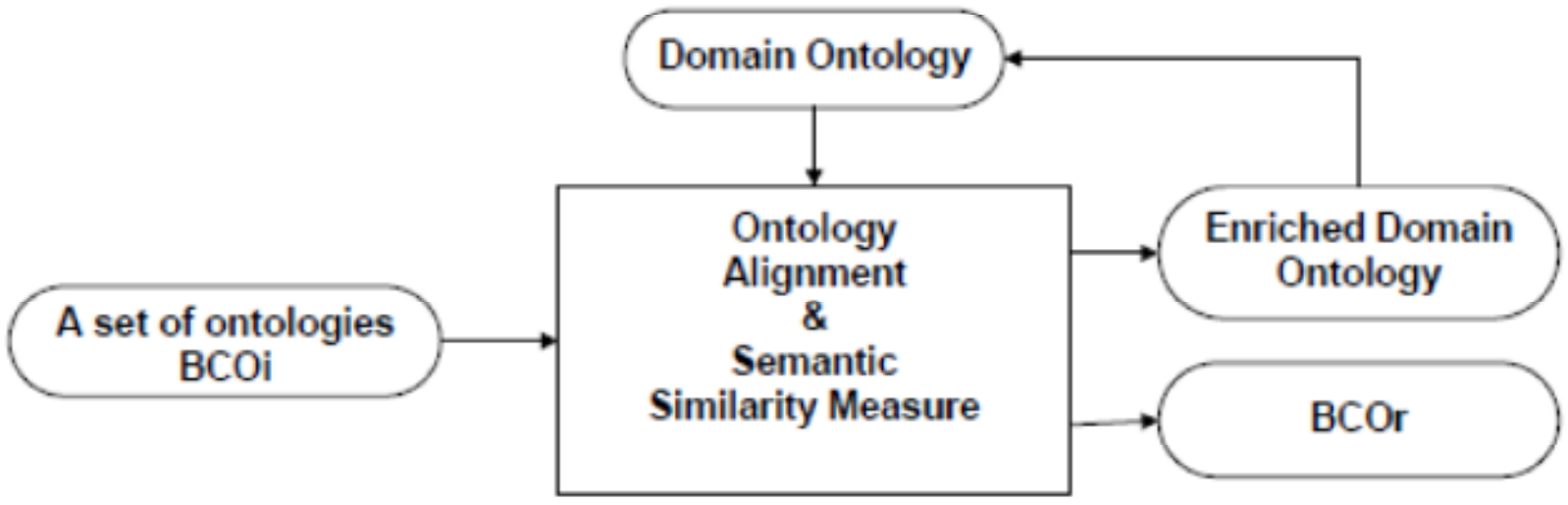

Figure 3. Each BCi to integrate, is transformed into an ontology BCOi [6]

*B. Ontologies alignement.*

This step consists in aligning ontologies obtained from the transformation of BC. We can use any alignment method based on targeted complementary resources, also called background ontologies or support ontologies [21,] [16], [23] and [6]. The domain ontology plays the role of targeted complementary resource and thus will be the support of ontologies alignment. This step of the process takes as input:

- A set of ontologies corresponding to each BC to alignment. These ontologies, denoted (BCOi) in figure 3, are outputted from the last process.
- The domain ontology chosen to support the alignment.

The outputs of this process are:

- Ontology, denoted BCOr in figure 3, resulting from the alignment of all BCOi ontologies

*C. Production of the correspondence ontology among BC*

Alignment process of ontologies derived from BC candidates for integration. This process gives an output a Correspondence Ontology (CO) among the concepts of BC. Based on CO among concepts to product another Correspondence Ontology among BC (BCCO), which will later be used either as external resources or support in the semantic integration process is to support IS designers to achieve their design tasks. Each type of relationship can highlight a conflict is syntactic, semantic or structural.

### 3.2.2. Semantic integration process

The inputs of the integration process are:

- A set of business components, denoted BC1 ... BCn, selected by the designer for inclusion in the future IS.
- Correspondence Ontology among BC (BCCO) result of pre-integration process.
- A catalogue of conflict resolution rules and integration rules which includes a set of resolutions rules (for example resolution rule of homonymy conflict is the re-naming by different names).

At this stage of integration, correspondence ontology can be exploited in various ways:

Table 1. The semantic relations and actions derived.

| | Semantic relation type in ontology | Actions proposed designers |
|---|---|---|
| The two concepts belong to correspondence ontology | Synonymy relation | Rename by the same name |
| | homonymy relation | Rename by different names |

The integration process outputs:

- A new Business Component result of the integration of a set of the BC.

The output of the process can be used later in future integrations for new components: The new Business Component result can be used as a candidate for integration with other components.

A. *Production of BC result.*

To demonstrate how to use correspondence ontology, we present resolution rules for naming conflicts derived from semantic relation: homonym and synonym existing in correspondence ontology result of our process.

***Conflict Resolution Rule 1:*** *if we have a semantic relation type synonym in the correspondence ontology between concepts of sources ontologies, we offer IS designer to rename the concepts with same name.*

***Conflict Resolution Rule 2:*** *if we have a semantic relationship type homonym in the correspondence ontology between concepts of sources ontologies, we offer IS designer to rename the concepts with different names.*

The figure 4 below shows a namely conflict resolution assisted by IS designer based on a set of conflict resolution rules stored in a catalogue. Based on correspondence ontology and the conflict resolution rules, we offer IS designer a decisions set represented by derived operations set. For example, if exist a relationship type synonym in correspondence ontology then find in the catalogue the resolving conflicts (conflict resolution rule 1), then propose to IS designer an operation “rename” one of concepts in conflicts and merge the two concepts or delete one of the concepts.

B. *Semantic integration of business components assisted by the designer of the information system*

In this section we present an overview of the chronological steps of semantic integration process of business components and how the information system designers can exploit the output of this process:

Chronological stages of the semantics integration process of BC are detailed below:

1. Returns the Semantic Relationships (SR) between concepts from business components candidates defined in correspondence ontology.
2. Find SR in the catalogue of rules for resolving conflicts.
3. Get to the rules of conflict resolution associated to SR.
4. Propose actions associated with these rules for resolving conflicts to information system designer.
5. The designer executes the default actions; he may choose other actions depending on the context.
6. Store actions chosen by designers for use later.
7. Produce a component result.

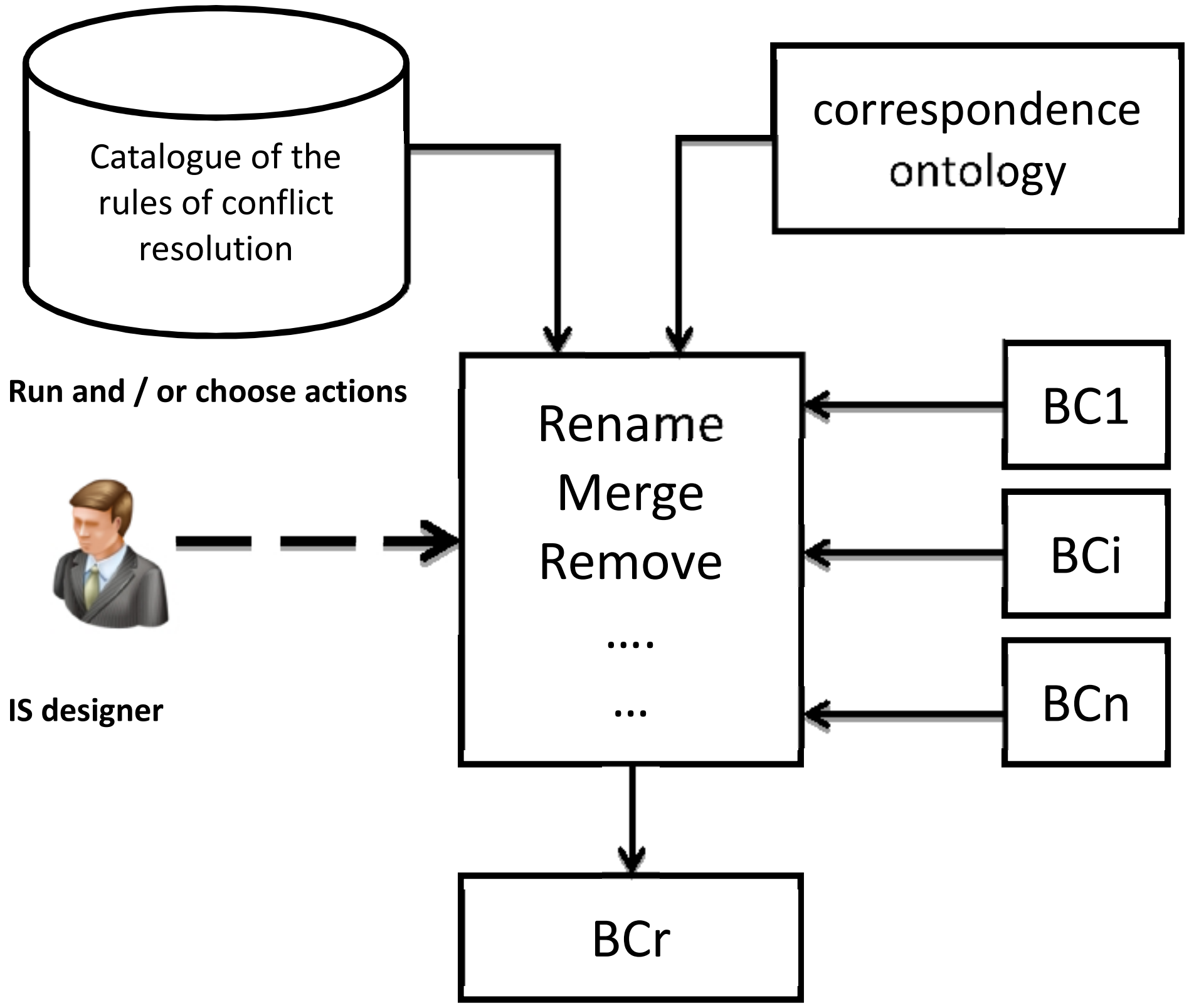

Figure 4. Illustrates the semantic integration process of business components.

### 3.3. Using Actions designers to optimize the integration process

The aim of this section is to provide a track for improving the relevance of semantic integration process of business components. The idea is to store choice for designers exploited in future treatments. The choices are represented by the pair (Relation, Action) they are stored in a database, which is designed to receive and retain the choice of designers. In our opinion this storage presents many advantages. In fact besides being able to keep the choice of designers and their contexts, such a basis can increase and improve the relevance of our semantic integration. Our process can be based on these experiences stored in each new task integration performed by the designers. if the designers chose an action A with a frequency N for a relation R exists in correspondence ontology. If N> THRESHOLD for Relation R (The THRESHOLD can be set by the user) we offer designers in the same context the same action.

For example for a synonymy relation in the correspondence ontology, the default action proposed is to rename the designer synonymous concepts by the same names, but the designer has chosen action "merge concepts" N times (by example N> 2) in the same context. therefore in future iterations we propose action "merge concepts" for this context.

## 4. ILLUSTRATIONSAND VALIDATION PROTOTYPE.

### 4.1. Example

In order to validate our proposal, we give an example followed by a prototype which we have developed. We illustrate the integration process using an example based on a real domain ontology called (The SWRC Ontology - Semantic Web for Research Communities) (Figure 6) and two components (Figure 5) related to "system management conferences." The ontology SWRC (Semantic Web for Research Communities) aims main modeling entities from Research Communities as individuals, organizations, publications (bibliographic meta-data), research topic and their relationships [29], ontology is available for download on the web link (http://ontoware.org/swrc/swrc/SWRCOWL/swrc_updated_v0.7.1.owl). We have extended this ontology with new concepts and relationships, this enriched ontology serves as ontology to support the process of semantic integration, it is represented and explored by the tool Protege1 and in particular plugin Jambalaya2. The business components denoted BC1 and BC2, described in UML represent the candidate components to semantic integration.

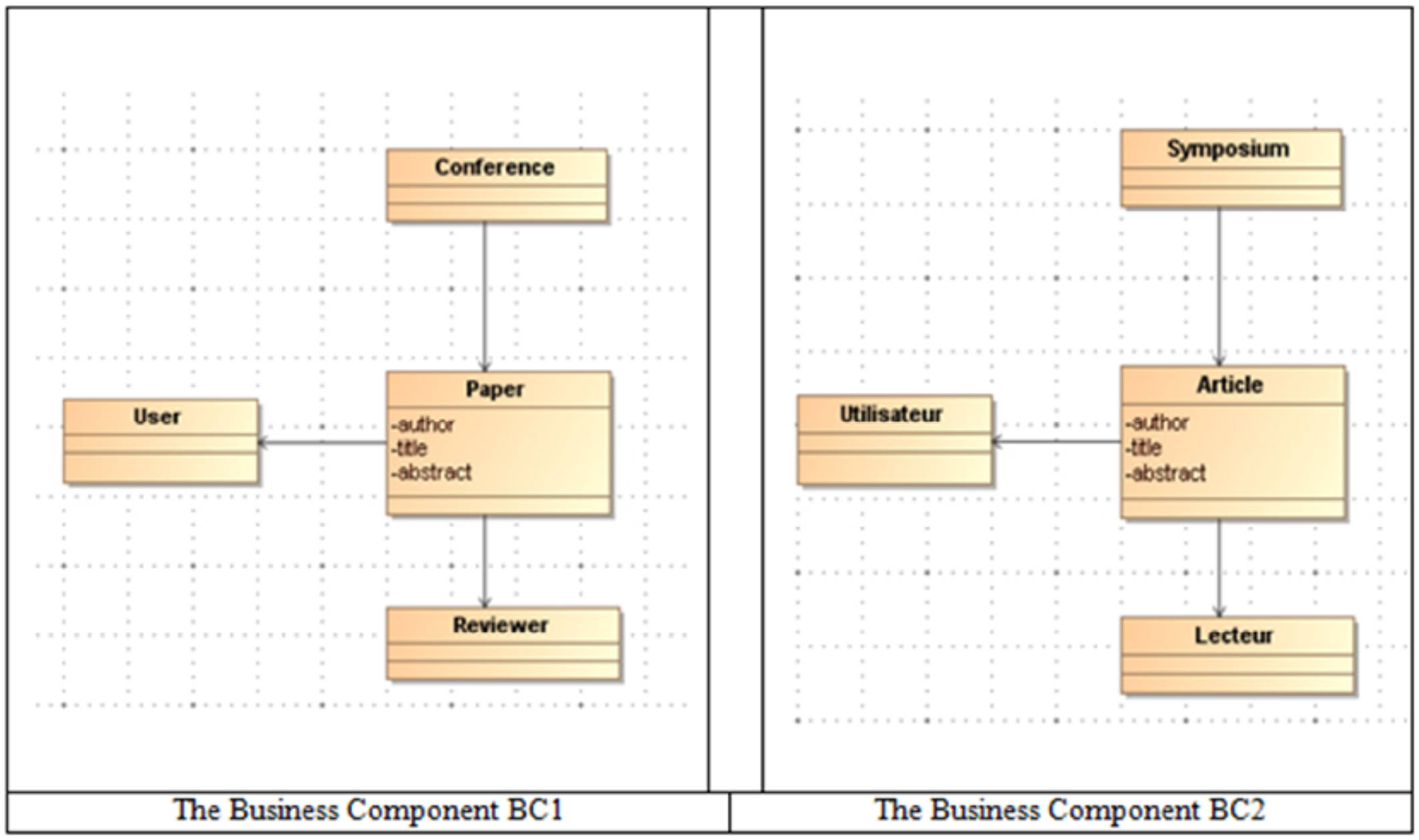

Figure 5: The two Business Component BC1 and BC2 candidates to integration

The figure below shows the hierarchy of the concepts of the domain ontology generated by the plugin Protege "Jambalaya."

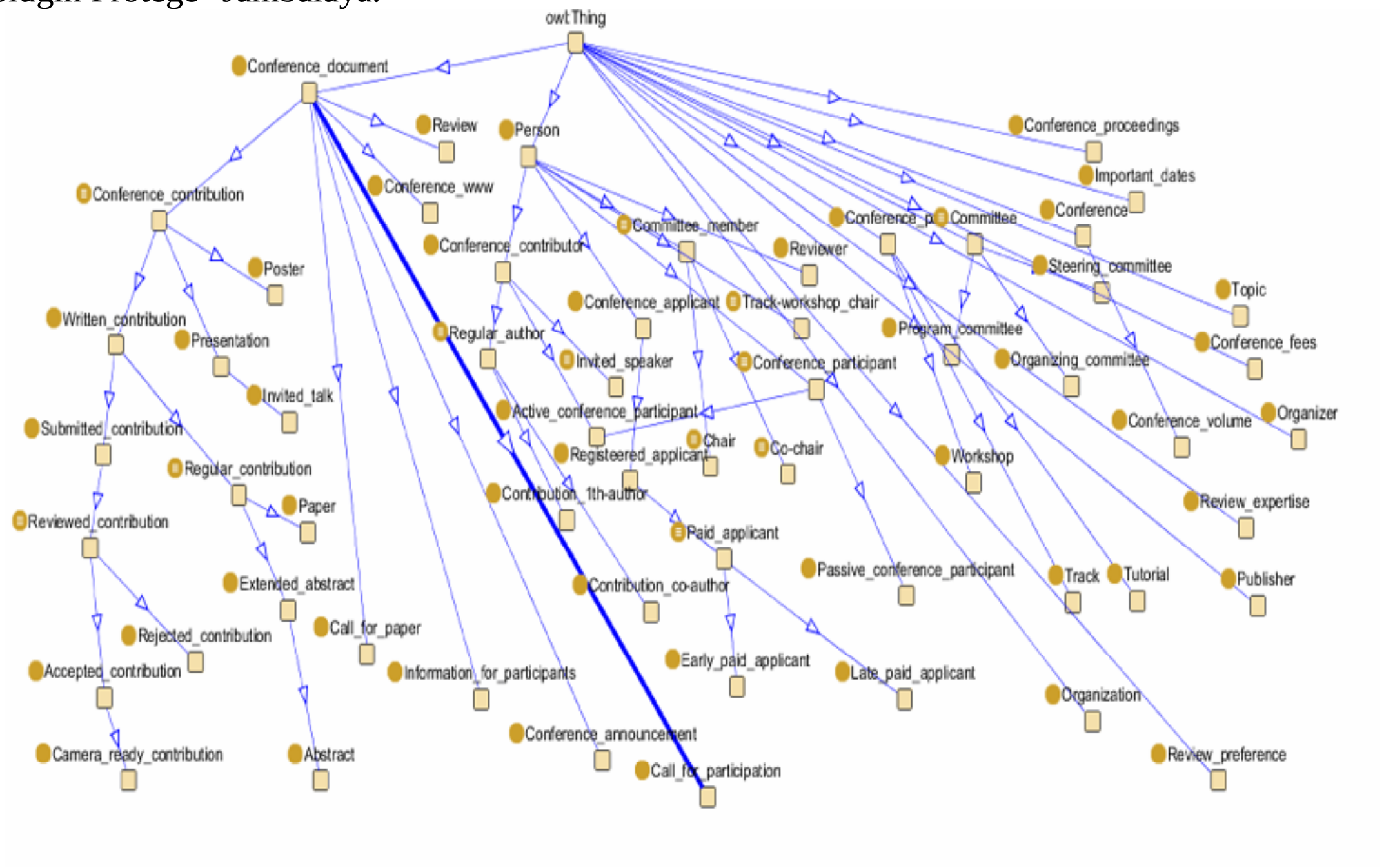

Figure 6. The domain ontology

Step 1: Transformation of BC1 and BC2 into ontologies. We transform the BC1 (BC2 respectively) to OBC1 (resp to OBC2).

Step 2: Alignment and obtaining semantic correspondence ontology with highlighting enrichment.

The ontology OBC1 derived from the component BC1 uses a concept called "Paper". Ontology OBC2 derived from the component BC2 uses a concept called "Article". The two concepts are in the domain ontology (C1 and C2 ∈ OD) and without allowing semantic relation between them (R (C1, C2) =∅.

The alignment of the two concepts requires therefore deduce the relationship through relationships of their sub-concept. Both concepts have sub-concepts “title", "abstract" and "author" that are similar. We deduce that "Paper" and "Article" are synonymous. Ditto for the concepts: "Conference", "User" and "Reviewer" from the component BC1 and concepts : "Symposium", "Lecteur" and "Utilisateur" from the component BC2. We deduce that "Conference" and "Symposium" are synonymous, "Lecteur" and "Reviewer" are synonymous and "User" and "Utilisateur" are synonymous. All these concepts and their relationships are added to the correspondence ontology.

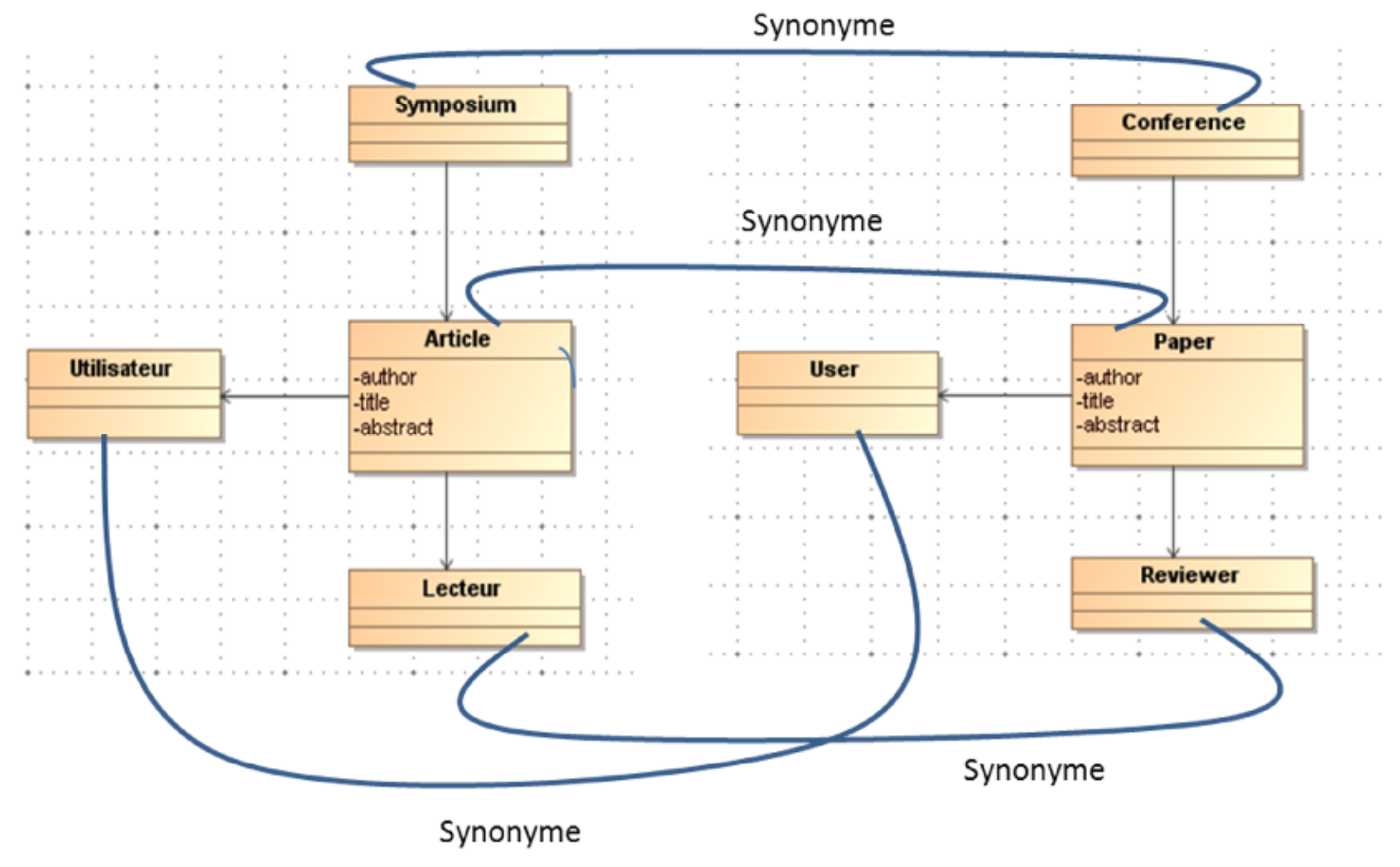

correspondence ontology

Figure 7. correspondence ontology

Step 3: The process of semantic integration
The integration process taking as input two components, correspondence ontology between the concepts of BC resulting from the pre-integration and a catalog of rules conflict resolution, which comprises a set of rules for a resolution. Information system designers can rely directly on the correspondence ontology and deduce that BC1 and BC2 are synonymous and that BCR is one of the two.

Based on the correspondence ontology and a catalog of rules of conflict resolution, we can offer designers the concepts in conflict and their relationship type, in our case the concepts are synonymous and actions to apply in this case is rename one of the concepts by the name of the other, or combine the two concepts.

### 4.2. **Prototype**

The last step of our work is to developing a prototype not only to validate and evaluate our semantic integration process but also to have a framework that can be used for semantic integration of BC. We describe in this section our prototype for the integration of BC and based on the integration process presented in the previous section.

The purpose of this prototype is to provide an interface for the user especially designers to achieve integration through semantic alignment of ontologies from BC by establishing correspondences between ontologies entities concerned. This correspondence will deduce the rules of integration and then starts execution.

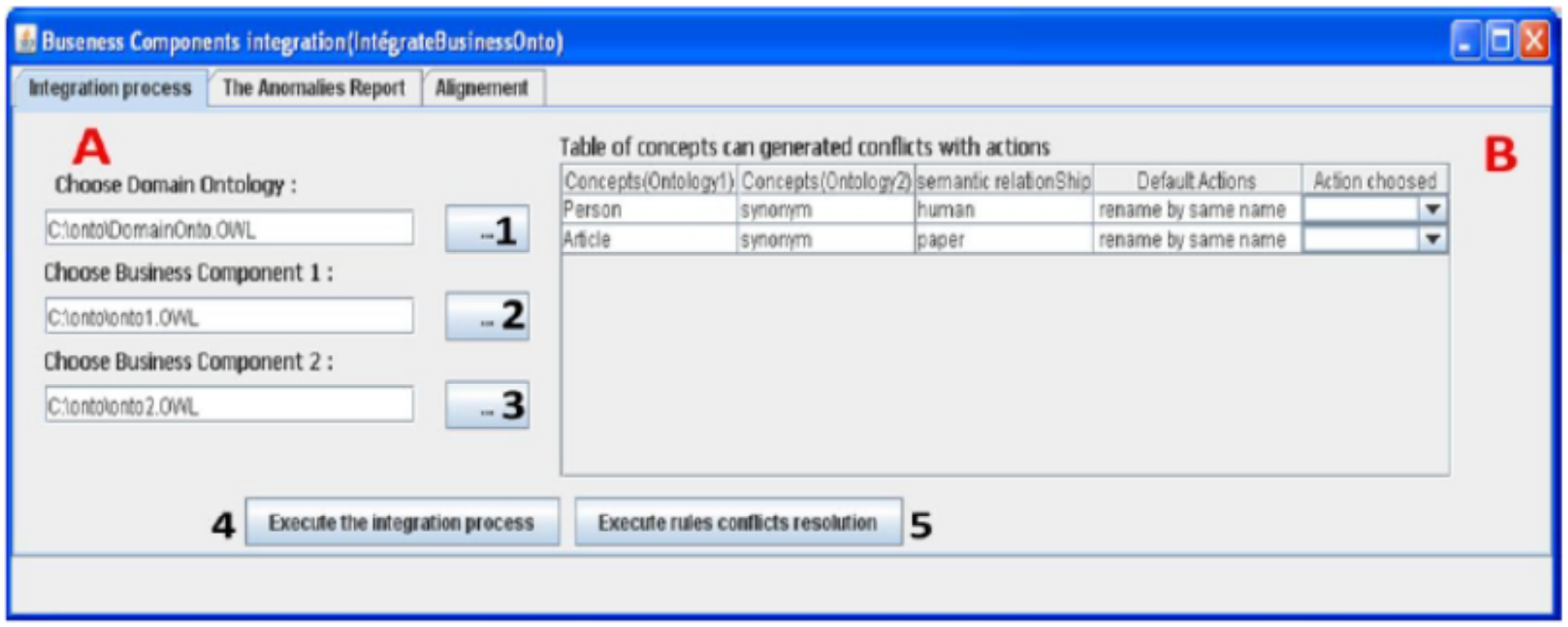

Figure 8. Schematic of the prototype GUI

Our prototype also provides the user the ability to view the XML code that related to the correspondence ontology result of the alignment. Show in the previous figure that pressing the button "Execute the integration process" will display the XML code of the correspondence ontology mapping in the "Alignment" tab.

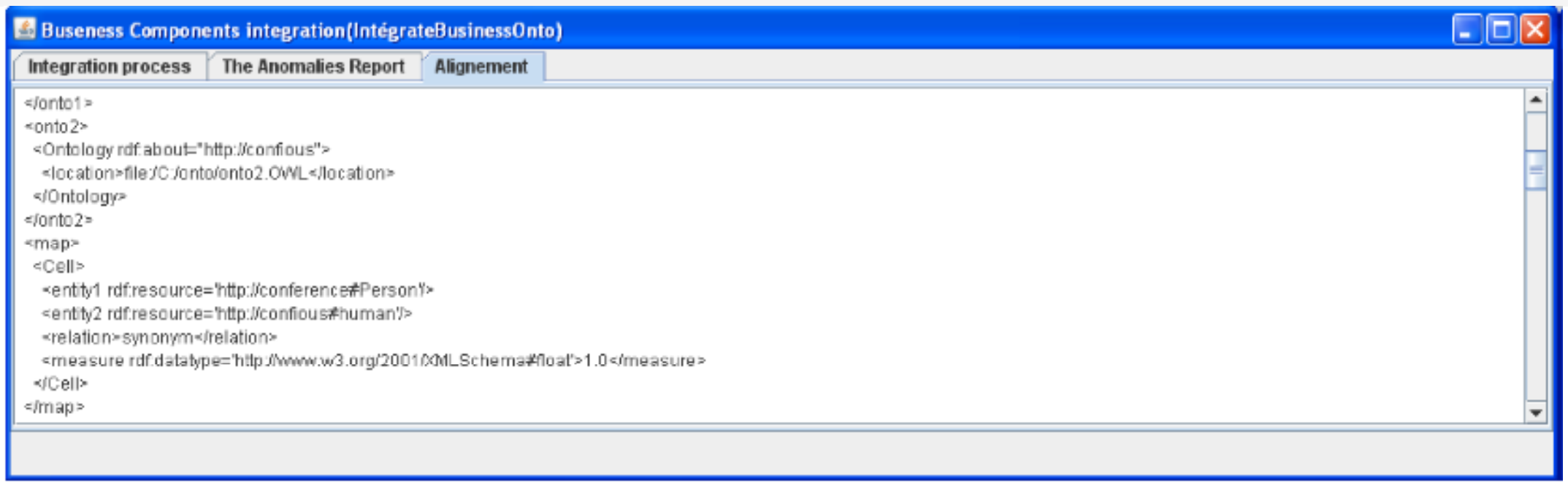

Figure 9. Correspondnace ontology in XML format

## 5. DISCUSSION

We believe that adding a preliminary step in our process of semantic integration is appropriate, including a postcondition step before preintegration step. Step postcondition is to improve the consistency of our integration process, using approaches Subgraph Isomorphism Search, Group isomorphism, and the check the existing of an isomorphic between two ontologies or BC that can eliminate many business components candidates for integration who does not respect the rules of an isomorphic by applying some mathematical elements: degree a node, the cardinal of a set …

***Isomorphic*** *”Two graphs are isomorphic if there is a one-to-one correspondence between their vertices and there is an edge between two vertices of one graph if and only if there is an edge between the two corresponding vertices in the other graph”[30].*

We hypothesize that two BC1 and BC2 can combine or integrate them as subBC1 □ BC1 and subBC2□ BC2 if and only if there exists a isomorphism f between subBC1and subBC2 as

f : subBC1→ subBC2 then subBC1□ subBC2

So two business components can be integrated if hypothesis above is verified the, that mean verify the existence of an isomorphic between the BC, this last well-known difficult and complex is known to be NP-complete and known in the literature by The Group Isomorphism problem (GIP) which analyzed in [31], in this sense we think proposing rules or algorithms for verify or test the non- isomorphic between the BC to deduce the relationships between them in the lover steps. In order to show the feasibility of this proposal, we propose the following rules.
We defined two operations SetType and Size (SetType)

- SetType an operation that returns a finite set of external relations syntactic, structural and semantic links a subBCi with other components.
- Size (Set) an operation that returns the number of elements in a set and size (SetType) represent the degree in graph theory.

*f between subBC1 and subBC2 is non-isomorphic if and only if (A) or (B) are satisfied* :

A. SetType(subBC1)≠SetType(subBC2)
B. size (SetType(subBC1))≠ size(SetType(subBC2))

## CONCLUSIONS

Our research is part of engineering information systems for reuse. We are interested specifically in conflict resolution type semantics in naming the reuse of business components in the conceptual phases of analysis and design. Our solution is based on an application of ontologies and their alignments IS design by reuse of conceptual business components. Application examples helped illustrate our approach.

We expect to continue to research the possibilities of expanding to solve other semantic conflicts, including conflicts measurement and confusion. The most important work was a process of semantic integration to support designers SI.

## REFERENCES

[1] Barbier F., Atkinson C., "Business Components", Business Component-Based Software Engineering, Kluwer, vol. 705, Chap. 1, pp. 1-26, 2002.

[2] R. Saidi, « conception et usage des composants métier processus pour les systèmes d'information », thèse de doctorat à l'institut polytechnique de Grenoble, le 26 Septembre 2009.

[3] I. Hassine, D. Rieu, F. Bounaas, O. Seghrouchni: Symphony: un modèle conceptuel de composants métier. INFORSID 2002: 167-182.

[4] V. Pujalte, P. Ramadour, C. Cauvet, "Recherche de composants réutilisables : une approche centrée sur l'assistance à l'utilisateur", in : Actes du 22ème congrès Inforsid, Biarritz, pp. 211-227, 25-28 mai 2004.

[5] Kzaz L., Elasri H., Sekkaki A., « A Model for Integration of Business Components ,International Journal of Computer Science & Information Technology (IJCSIT), Volume II, Number 1 pp 1-12, ISSN:0975-3826 (Online); 0975-4660 (Print), February 2010.

[6] Hicham Elasri, Abderrahim Sekkaki, Larbi Kzaz: An Ontology-Based Method for Semantic Integration of Business Components, IEEE New Technologies of Distributed Systems (NOTERE), 2011 11th Annual International Conference on, 9-13 May 2011, Paris, france.

[7] Larbi Kzaz, Hicham Elasri, Abderrahim Sekkaki, Résolution des conflits sémantiques pour l'intégration des composants métier, 4èmes Journées Francophones sur les Ontologies - 22 – 23 Juin 2011, Montréal, Canada

[8] Stefan Seedorf, Martin Schader: Towards an Enterprise Software Component Ontology. Americas Conference on Information Systems (AMCIS) 2011: Detroit, Michigan, USA

[9] Herzum, P. and Sims, O. : Business component factory: A comprehensive overview of component-based development for the enterprise. Wiley, New York, 2000.

[10] Philippe Ramadour, Corine Cauvet: Approach and Model for Business Components Specification. Database and Expert Systems Applications (DEXA), Aix-en-Provence, France, 2002: 628-637

[11] Colin Atkinson, Thomas Kühne, The role of metamodeling in MDA , In Proc. UML 2002 Workshop Software Model, pp. 67-70, Dresden, Germany, October 2002

[12] Jean Bézivin, Salim Bouzitouna, Marcos Didonet Del Fabro, Marie-Pierre Gervais, Frédéric Jouault, Dimitrios S. Kolovos, Ivan Kurtev, Richard F. Paige: A Canonical Scheme for Model Composition, ECMDA-FA 2006: Bilbao, Spain

[13] P. Lando, « Conception et développement ’applications informatiques utilisant des ontologies : application aux EIAH » 1res Rencontres jeunes chercheurs en EIAH, RJC-EIAH,2006.

[14] Gruber, T. What is an ontology? Retrieved November, 2006, from http://www-ksl.stanford. edu/kst/what-is-an-ontology.html,

[15] Viviana Mascardi, Angela Locoro, Paolo Rosso: Automatic Ontology Matching via Upper Ontologies: A Systematic Evaluation. IEEE Trans. Knowl. Data Eng. 22(5): 609-623, 2010

[16] Safar B, Chantal R., « Alignement d’ontologies basé sur des ressources complémentaires : illustration sur le système TaxoMap » In Revue TSI, 22p, 2009

[17] Shvaiko P., Euzenat J., A survey of schema-based macthing approaches. Journal on Data Semantics (JoDS), IV, 2005.

[18] Aleksovski Z., Klein M., Ten Kate W., Van Harmelen F. «Matching Unstructured Vocabularies using a Background Ontology », Proceedings of the 15th International Conference on Knowledgee Springer-Verlag, p. 182-197.,2006

[19] W. R. van Hage, S. Katrenko, and G. Schreiber. A method to combine linguistic ontology-mapping techniques. In Yolanda Gil, Enrico Motta,V. Richard Benjamins, and Mark A. Musen, editors, Proc. of ISWC 2005, pages 732–744. Springer, 2005.

[20] J. Gracia, V. Lopez, M. d’Aquin, M. Sabou, E. Motta, and E. Mena. Solving semantic ambiguity to improve semantic web based ontology matching. In P. Shvaiko, J. Euzenat, F. Giunchiglia, and B. He, editors, Proc. of OM-2007, 2007.

[21] M. Sabou, M. d’Aquin, and E. Motta. Using the semantic web as background knowledge for ontology mapping. In P. Shvaiko, J. Euzenat, N. Noy, H. Stuckenschmidt, R. Benjamins, and M. Uschold, editors, Proc. of OM-2006, 2006.

[22] C. Faucher, F. Bertrand J.Y. ,2006 : Génération d’ontologie à partir d’un modèle métier UML annoté, REVUE DES NOUVELLES TECHNOLOGIES DE L'INFORMATION E, 12 (2008) 65

[23] Gašević, D., Djurić, D, Devedžić, V., Damjanović, V., —UML for Read-To-Use OWL Ontologies, In Proceedings of the IEEE International Conference Intelligent Systems , Vrana, Bulgaria, 2004.

[24] Zedlitz, J., Jorke, J., Luttenberger, N.: From UML to OWL 2. In: Proceedings of Knowledge Technology Week 2011. Springer,2012

[25] Atkinson, Gutheil, and Kiko: On the Relationship of Ontologies and Models. In:Proceedings of the 2nd Workshop on Meta-Modeling and Ontologies, p. 47{60. Gesellschaft fur Informatik, Bonn, October 2006.

[26] Hart, Emery, Colomb, Raymond, Taraporewalla, Chang, Ye, Kendall, and Dutra: OWL Full and UML 2.0 Compared, March 2004.

[27] Kiko, and Atkinson: A Detailed Comparison of UML and OWL. Technischer Bericht 4, Dep. for Mathematics and C.S., University of Mannheim, 2008.

[28] Falkovych, K. Sabou, M. Stuckenschmidt, H. UML for the Semantic Web:Transformation-Based Approaches in B. Omelayenko and M. Klein: Knowledge Transformation for the Semantic Web, IOS Press, 2003

[29] York Sure, Stephan Bloehdorn, Peter Haase, Jens Hartmann, Daniel Oberle: The SWRC Ontology - Semantic Web for Research Communities. Portuguese Conference on Artificial Intelligence (EPIA) 2005: 218-231

[30] Seth A. Greenblatt. Ontologies for graph matching: Practice and potential. From a Presentation at Workshop on Associating Semantics With Graphs, April 16{17 2007. http://dydan.rutgers.edu/Workshops/Semantics/slides/greenblatt.pdf.

[31] Youming Qiao, Jayalal Sarma M. N., Bangsheng Tang: On Isomorphism Testing of Groups with Normal Hall Subgroups. Journal of Computer Science and Technology, Volume 27 (4): 687-701 ,2012

**Authors**

**Hicham ELASRI** is currently a doctoral student in Faculty of Science, University Hassan II Aïn Chock Morocco. He does research on semantic interoperability of distributed information system and Geographic Information System (GIS)

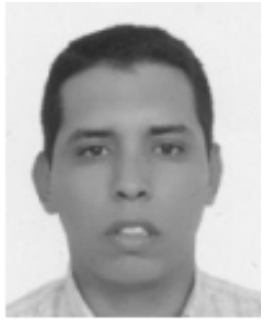

**Abderrahim Sekkaki** received a D.Sc. in Network Management domain from the Paul Sabatier University, France, in 1991: and a Dr. of State Degree from Hassan II University, Morocco, in 2002. He does research on distributed systems and policies based network management. Presently, he is a Professor in Computer Science at the Hassan II University, Casablanca, Morocco

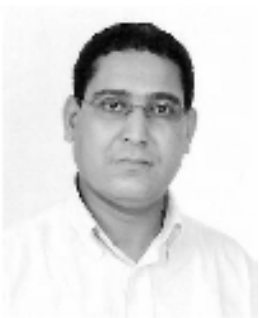